\documentclass[12pt]{article}
\usepackage{graphicx}
\begin{document}
\baselineskip=5.5mm
\newcommand{\be} {\begin{equation}}
\newcommand{\ee} {\end{equation}}
\newcommand{\Be} {\begin{eqnarray}}
\newcommand{\Ee} {\end{eqnarray}}
\def\lg{\langle}
\def\rg{\rangle}
\def\a{\alpha}
\def\b{\beta}
\def\g{\gamma}
\def\G{\Gamma}
\def\d{\delta}
\def\D{\Delta}
\def\eps{\epsilon}
\def\k{\kappa}
\def\l{\lambda}
\def\L{\Lambda}
\def\om{\omega}
\def\Om{\Omega}
\def\t{\tau}
\noindent
\noindent
\begin{center}
{\Large
{\bf
Kerr effect as a tool for the investigation of dynamic heterogeneities
}}\\
\vspace{0.8cm}
{\bf Uli H\"aberle and Gregor Diezemann} \\

Institut f\"ur Physikalische Chemie, Universit\"at Mainz,
Welderweg 11, 55099 Mainz, Germany
\\

\end{center}
\vspace{1cm}
\noindent
{\it
We propose a dynamic Kerr effect experiment for the distinction between dynamic heterogeneous and homogeneous relaxation in glassy systems. The possibility of this distinction is due to the inherent nonlinearity of the Kerr effect signal. We model the slow reorientational molecular motion in supercooled liquids in terms of non-inertial rotational diffusion. The Kerr effect response, consisting of two terms, is calculated for heterogeneous and for homogeneous variants of the stochastic model. It turns out that the experiment is able to distinguish between the two scenarios. We furthermore show that exchange between relatively 'slow' and 'fast' environments does not affect the possibility of frequency-selective modifications. It is demonstrated how information about changes in the width of the relaxation time distribution can be obtained from experimental results. 
}

\vspace{0.5cm}

\noindent

\vspace{1cm}
\noindent

\section*{I. Introduction}

Correlation functions in disordered systems usually decay nonexponentially in time \cite{angell}. The microscopic origin of this behavior is of importance for a detailed understanding of the dynamics in these systems. In principle, the stretched exponential form of the correlation functions can be explained by two extreme scenarios. Either different Debye-type relaxation rates superimpose to the total response, or the broadened form is an intrinsic property of the dynamics in glassy systems. These different scenarios have been denoted as dynamic heterogeneous and homogeneous \cite{hethom}. Several, but not numerous, experimental techniques have been developed that allow the investigation of dynamic heterogeneity, like four-time NMR measurements \cite{4timenmr}, a deep bleach experiment \cite{deepbleach} and nonresonant hole burning (NHB) \cite{NHB}. 

In a theoretical investigation of NHB \cite{gregor} it has been shown that indeed heterogeneous and homogeneous dynamics can be distinguished. Experimental realizations of NHB have been conducted on numerous systems like supercooled liquids \cite{NHB,duvvuri,blochi}, relaxor materials \cite{oli} and spin glasses \cite{chamberlin}. In all these studies indications for dynamic heterogeneity surpass those for homogeneity, but also homogeneous behavior was found, e.g., in an amorphous ion conductor \cite{richert}. Recently a mechanical variant of NHB has been developed \cite{mckenna}.

Mainly accountable for the distinction between heterogeneous and homogeneous dynamics is the possibility to adress certain dynamics in the sample, i.e., the selection of specific dynamic subensembles. Frequency-selective behavior is obtained only if the measured response function is nonlinear in the applied field, which holds for the dynamic Kerr effect \cite{kerreffect} as well as for NHB. In Kerr effect relaxation, frequency-selective behavior can be achieved if a driving AC field with varying frequency is applied.  We demonstrated frequency-selectivity in a Kerr effect experiment in the terahertz range in a recent publication \cite{mykerr}. Treating the vibrational dynamics around the boson peak in a Brownian oscillator model we concluded that the frequency dependence of the vibrations' damping can be extracted experimentally. 

The Kerr effect has to our knowledge not yet been used for an investigation of dynamic heterogeneities. In this article, we study the Kerr effect response in the range of slow reorientational motions in supercooled liquids. We propose the Kerr effect for a distinction between dynamic heterogeneous and homogeneous relaxation. The advantage of the Kerr effect compared to NHB lies in the fact that the Kerr effect response is purely nonlinear in the applied field. Therefore no separation of nonlinear contributions from the linear response background is required, and field strengths remarkably weaker than those usually applied in NHB ($100\ kV/cm$ \cite{NHB,duvvuri,blochi}) may suffice. It should also be possible to study dynamics at different temperatures ranging from the glass transition temperature to high temperatures. Here, possible changes in a heterogeneous distribution with temperature \cite{breitetemp,gammaverteilung} might be investigated.  

The paper is organized as follows. Section II outlines the principles of the Kerr effect, explains our suggested experiment and defines heterogeneous and homogeneous models. In Section III we give expressions for the Kerr effect response in the different approaches and we discuss the results. It turns out that indeed the distinction between dynamic homogeneity and heterogeneity is possible. The conclusions are given in Section IV.

\section*{II. Theory}
\subsection*{1. Dynamic Kerr effect}
If an anisotropically polarizable sample is exposed to a time-dependent electric field it becomes birefringent. This phenomenon is known as dynamic Kerr effect. In a theoretical description the coupling of external electric fields $E(t)$ to matter via the sample's permanent dipole moment $\mu$ has to be taken into account as well as the coupling via the polarizability $\alpha$. The structure of the Hamiltonian describing the interaction is thus of the form \cite{hamop}
\be \label{exthamilton}
\mathcal{H}_{ext}=-\mu E(t)P_1(\cos \theta)-\alpha E^2(t) P_2(\cos \theta)
\ee
The permanent dipole moment interaction is linear in $E$, while the polarizability interaction is of order $E^2$ since the induced dipole moment itself is $\propto E$. The appearance of the Legendre polynomials $P_L(\cos \theta)$ is due to the tensorial nature of the dipole moments. Since the permanent dipole moment is a vector and the polarizability is a matrix, they transform like first and second rank Legendre polynomials, respectively. The scalar $\alpha$ in Eq.(\ref{exthamilton}) is to be identified with the difference between the polarizabilities parallel and perpendicular to the internal symmetry axis \cite{deschardaeng}, and $\theta$ is the angle between this axis and the applied field. 

The time-dependent polarizability is the quantity of interest for a description of the dynamic Kerr effect. We therefore have to calculate the expectation value of the second Legendre polynomial
\be\label{alpha}
\lg \alpha P_2(\cos \theta (t) )\rg  =\alpha \lg P_2(t) \rg .
\ee
The brackets denote an expectation value over the whole sample. In this work we focus on reorientational motions. Here, the time dependence of the orientation (described by the angle $\theta$) determines the time dependence of the polarizability. In the following, we will use the short hand notation on the right hand side of Eq.(\ref{alpha}).

The expectation value $\lg P_2(t) \rg$ is calculated in some dynamic model, where the applied external field that determines the time dependence is treated in perturbation theory. Even without specifying the model yet, it is clear that the linear response of order $E$ must vanish as long as isotropic systems are considered. This is because the linear response is always proportional to the first rank Legendre polynomial, see Eq.(\ref{exthamilton}). Because of the orthogonality of the Legendre polynomials the linear response vanishes if the expectation value of the second rank Legendre polynomial, Eq.(\ref{alpha}), is calculated. The Kerr effect response is thus of order $E^2$.
 
In some more detail we have two contributions to the expectation value $\lg P_2(t) \rg$. The first one is proportional to the polarizability $\alpha$ (formally calculated with the second term in the external Hamiltonian (\ref{exthamilton}) in first order perturbation theory), the second one is quadratic in the permanent dipole moment $\mu$ (second order perturbation theory with the first term in (\ref{exthamilton})). We will denote these contributions in the following as $\lg P_2^\alpha(t) \rg$ and $\lg P_2^{\mu \mu}(t) \rg$. 

\subsection*{2. Experiment}

We propose an experiment as follows. First, a sinusoidal electric field is applied for an arbitrary number $N$ of (full) cycles to a sample in equilibrium.
\be\label{feld}
E(\tau )=E \sin(\Omega_p \tau) \qquad \tau <t_p=\frac{2\pi N}{\Omega_p}
\ee
The most important experimental parameter is the pump frequency $\Omega_p$. After the 'pump time' $t_p$ the field is switched off and the relaxation of the sample's polarizability back to thermal equilibrium is measured. In our terminology the time variable $t$ starts with $0$ at $t_p$. 

The experiment is based on the same idea as NHB. If the dynamics is assumed to be heterogeneous (different relaxation rates), then it should be possible to address 'slow' or 'fast' dynamics separately by varying the pump frequency from 'low' to 'high'. The so-selected ensembles will then show 'slow' or 'fast' relaxation behavior afterwards. In a Fourier transformed representation of the relaxation, one would therefore expect an extremum position proportional to the applied field frequency. On the other hand, in a homogeneous system one would expect relaxation on a single time scale only, and different pump frequencies should not alter the time-dependence or extremum position in the Fourier transform. Thus, shifts in the extremum position can be considered as indications for dynamic heterogeneous relaxation while no shifts would support the homogeneous picture.

The aim of this paper is to demonstrate that the concept outlined above indeed holds in heterogeneous and homogeneous models. Because of the nonlinearity of the Kerr effect the contributions $\lg P_2^\alpha(t) \rg$ and $\lg P_2^{\mu \mu}(t) \rg$ are not trivially related to the correlation function, and have thus to be explicitly calculated for a given field sequence like Eq.(\ref{feld}). We specify the dynamic models used in the calculation of the response functions in the following section.

\subsection*{3. Dynamics in the rotational diffusion model}

For a description of the dynamics we use the well-known rotational diffusion model \cite{kubo}. The rotational diffusion equation for the conditional probability $P_i(\Om,t|\Om ',t')$ to find the orientation $\Om$ at time $t$, assuming it was $\Om '$ at time $t'<t$ is
\be\label{simplefopl}
\frac{\partial}{\partial t} P_i(\Om,t|\Om ',t')=\hat \Pi_i(\Om ,t) P_i(\Om,t|\Om ',t')
\ee
with the Fokker-Planck operator
\be\label{rotdiff}
\hat \Pi_i(\Om ,t) =D_i(t)\left[ \frac{1}{\sin\theta} \partial_\theta \sin\theta \left( \partial_\theta +\beta \frac{\partial {\mathcal H}_{ext}}{\partial \theta}\right) + \frac{1}{\sin^2\theta} \partial^2_\phi \right] .
\ee
Here, $\theta$ and $\phi$ denote polar and azimuth angles with respect to the molecules internal symmetry axis. These angles constitute the orientation $\Omega$. $D_i(t)$ is a (possibly time-dependent) rotational diffusion constant, $\beta=(k_B T)^{-1}$ and ${\mathcal H}_{ext}$ is the Hamiltonian for the interaction with the applied external field in Eq.(\ref{exthamilton}). 

The solution of Eq.(\ref{rotdiff}) without driving field (${\mathcal H}_{ext}=0$) is achieved via an expansion in Wigner rotation matrices $D^{L}_{MN}$ \cite{wigner}. The ansatz
\begin{equation}\label{ansatz}
P_i(\Omega,t|\Omega ',t')=\sum_{LMN} \frac{2L+1}{8\pi^2} D^{L}_{MN}(\Omega) \left[ D^{L}_{MN}(\Omega ')\right]^\ast G^{(L)}_i(t,t')
\end{equation}
inserted in Eq.(\ref{simplefopl}) yields
\begin{equation}\label{glvont}
G^{(L)}_i(t,t')=\exp\left[ -L(L+1)\int_{t'}^t d\tau \ D_i(\tau)\right]
\end{equation}
and rotational correlation functions of the $L$-th Legendre polynomial are given by
\be\label{corr}
 C^L_i(t,t') = \lg  P_L(\cos\theta(t)) P_L(\cos\theta(t')) \rg_i =\frac{1}{2L+1} G^{(L)}_i(t,t'). 
\ee
In the presence of external fields perturbation theory is used to calculate the response functions. The corresponding perturbation operators result from the Fokker-Planck operator (\ref{rotdiff}) including the external Hamiltonian in Eq.(\ref{exthamilton}). Linear response functions obtained this way are related to the correlation function via the  fluctuation dissipation theorem \cite{calabrese}.

\subsection*{Heterogeneous model}
In a heterogeneous scenario one assumes the coexistence of different, 'slow' and 'fast' environments related to small and large rotational diffusion constants $D_i$. If $D_i$ is time-independent, the propagator $P_i(\Om,t|\Om ',t')$ is time-translational invariant and the correlation function in case of a single rotational diffusion constant is given by
\be\label{}
C_i^L(t,t')=C_i^L(t-t')=\frac{1}{2L+1} \exp\left[-L(L+1) D_i(t-t')\right].
\ee
Dynamic heterogeneities are known to have finite life times of similar magnitude as the $\alpha$-relaxation time \cite{otptauschtaus,austauschsim}. Finite life times can be modelled via exchange models. Various approaches exist to include exchange in a model \cite{xeigenwerte}. In a master equation ansatz \cite{sillescu} one assumes that slow environments become fast and vice versa. The environments are related to rotational diffusion constants, and exchange between the environments leads to effectively time-dependent rotational diffusion constants. Time-translational invariance is preserved within this framework. We will denote the environments by $\eps_i$ in the following, and we add a second index to quantities like the propagator. $P_{ij}(\Omega, t | \Omega ',t')=P_{ij}(\Omega, t-t' |\Omega ',0)$ thus is the probability to find the orientation $\Omega$ in environment $\eps_i$ at time $t$, assuming it was $\Omega '$ in environment $\eps_j$ at time $t'$. Transition rates $\gamma_{ij}$ determine transitions from environment $\eps_j$ to $\eps_i$, meaning that the rotational diffusion constant changes its value from $D_j=D(\eps_j)$ to $D_i=D(\eps_i)$. Furthermore, the behavior of the orientation during a transition has to be specified. If we assume that each transition is associated with a rotational random jump, then an exchange model can be defined by \cite{sillescu} 
\be\label{beckpfeif}
\frac{\partial}{\partial t} P_{ki}(\Om,t|\Om ',t')=\left( \hat \Pi_k(\Om)-\gamma_k \right)  P_{ki}(\Om,t|\Om ',t') +  \frac{1}{8\pi^2} \sum_{j \not= k} \gamma_{kj} \int d\Om_0 P_{ji}(\Om_0,t|\Om ',t').
\ee
Here, $\hat \Pi_k(\Om)$ is the Fokker-Planck operator (\ref{rotdiff}) with $D_i(t)$ replaced by the time-independent $D_k$ and $\gamma_k=\sum_{j\not= k}\gamma_{jk}$ is the total rate out of state $\eps_k$. 

The solution of Eq.(\ref{beckpfeif}) without external fields is obtained in a similar manner as the solution of Eq.(\ref{simplefopl}). Here, the function $G^{(L)}_i(t,t')$ in Eq.(\ref{ansatz}) is replaced $G^{(L)}_{ki}(t,t')$. For $L\not= 0$ the elements of the matrix ${\bf G}^{(L)}(t,t')$ are given by
\be\label{c0gdiag}
G^{(L\not= 0)}_{ki}(t,t')=\delta_{ki} \exp\left[ \Lambda^{(L)}_i (t-t')\right]
\ee
with the eigenvalues
\be\label{xeigen}
\Lambda^{(L)}_i=-L(L+1)D_i-\gamma_i.
\ee
The equilibrium probabilities $P_i^{eq}$ to find the system in environment $\eps_i$ in the presence of exchange (finite $\gamma_{ij}$) are obtained from the long-time limit of the propagator, $P_i^{eq}=\lim_{(t-t') \to \infty} \int d\Omega  P_{ik}(\Omega,t|\Omega ',t')=G^{(0)}_{ik}(\infty ,0)$, which guarantees detailed balance. Correlation functions in the heterogeneous model are then calculated as
\be\label{hetcorr}
C^L(t)=\sum_i C^L_i(t) P^{eq}_i=\frac{1}{2L+1}\sum_i \exp\left[\Lambda_i^{(L)}t\right] P^{eq}_i .
\ee
The larger the exchange rates are, the more they dominate the eigenvalues in comparison to the $L$-dependent term in Eq.(\ref{xeigen}). This also holds in other exchange models \cite{xeigenwerte}. Large exchange rates thus lead to a weaker $L$-dependence of the eigenvalues, and the exponents in correlation functions (\ref{hetcorr}) for different values of $L$ become more and more similar. The inclusion of exchange in the rotational diffusion model thus leads to very similar correlation times for different $L$, as usually found experimentally \cite{willi}. The calculation of response functions is straightforward.

Exchange effects are neglected if $\gamma_{ij}=0$ is chosen. Then the eigenvalues in Eq.(\ref{hetcorr}) are given by $\Lambda_i^{(L)}=-L(L+1)D_i$, and a distribution of rotational diffusion constants for a heterogeneous model can be arbitrarily chosen. If the relaxation times $\tau_i \propto D_i^{-1}$ are distributed according to the generalized Gamma distribution \cite{gammaverteilung}, a stretched exponential form of the correlation function is obtained.

\subsection*{Homogeneous model}
In a homogeneous scenario one assumes that the relaxation rate (or rotational diffusion constant) depends on time. A decaying power law function
\be\label{homd}
D_i(t)=\frac{\beta_K}{2\tau}  \left( t/\tau \right)^{\beta_K-1} 
\ee
inserted into the correlation function (\ref{corr}) using Eq.(\ref{glvont}) yields 
\be\label{homcorr}
\lg P_1(t)P_1(t')\rg = \frac{1}{3} \exp\left[ -(t'/\tau)^{\beta_K}  \right] \exp\left[ -(t/\tau)^{\beta_K}  \right] .
\ee
Thus, the special choice in Eq.(\ref{homd}) reproduces a stretched exponential for the correlation function. Note that Eq.(\ref{homcorr}) is not time-translational invariant, which is due to the time dependence of $D_i(t)$ \cite{calabrese}. Due to the lack of time-translational invariance the system has an 'internal clock' that is hard to justify from a physical point of view. The consequences in our case will be discussed below. An alternative for modelling homogeneous dynamics are the so-called fractional Fokker-Planck approaches \cite{fractional}. Here, time-translational invariance is maintained \cite{fracttti}.

\section*{III. Results}
In this section we discuss the Kerr effect response to the alternating field in the various approaches considered in the previous section. After the orientational integrations have been carried out we find the expressions 
\be\label{alal}
\lg P_2^{\alpha}(t) \rg=\frac{6}{5} \beta \alpha E^2 \sum_{i} \int_0^{t_p} dt' \sin^2(\Omega_p t') D_i(t') G_{i}^{(2)}(t+t_p,t') P_i^{eq}
\ee
for the $\alpha \alpha$- contribution to the Kerr effect response and
\Be \nonumber  
\lg P_2^{\mu \mu}(t) \rg=\frac{4}{5}\beta^2 \mu^2 E^2 \times \\ \label{almumu}
\times \sum_{i} \int_0^{t_p} dt' \int_0^{t'} dt'' \sin(\Omega_p t') \sin(\Omega_p t'') D_i(t') D_i(t'') G_{i}^{(2)}(t+t_p,t') G_{i}^{(1)}(t',t'') P_i^{eq} 
\Ee
for the $\alpha \mu \mu$-contribution. Eqns.(\ref{alal},\ref{almumu}) are valid for heterogeneous and homogeneous models. In the heterogeneous case the $D_i$ are time-independent and in the homogeneous case no summation over $i$ is required. In the presence of exchange $G_i^{(L)}(t_1,t_2)$ is to be identified with the diagonal elements of the matrix ${\mathbf G}^{(L)}(t_1,t_2)$, see Eq.(\ref{c0gdiag}).

Note that Eq.(\ref{alal}) is formally identical to the expression for the linear (dielectric) response, except for the fact that the external field of sine form enters quadratically and $G^{(2)}_i$ appears instead of $G^{(1)}_i$. As a consequence of this quasi-linearity the $\alpha \alpha$-contribution is -like the dielectric linear response- unable to distinguish between homogeneous and heterogeneous scenarios.

Let us briefly discuss the simplest case where the system is described by a single, time-independent rotational diffusion constant. Here, $G^{(L)}_i(t_1,t_2)$ are exponential functions according to Eq.(\ref{glvont}). We furthermore have to set $D_i=const$ and $\sum_i P^{eq}_i=1$ in Eqns.(\ref{alal},\ref{almumu}). The time integrals can then be carried out analytically, leading to response functions of the form
\be\label{respeinerallg}
\lg P_2^{\Theta}(t) \rg=A^{\Theta}(\Omega_p, D_i,t_p)\exp(-6 D_i t)
\ee
where $\Theta$ stands for $\alpha$ or for $\mu\mu$. In the following discussion of the response functions we will show the imaginary part of Fourier-transformed response functions, denoted as $\lg P_2^\Theta(\omega)\rg$. The symbol $\omega_{max}$ will be used for the extremum positions in $\lg P_2^\Theta(\omega)\rg$.

Both contributions (\ref{respeinerallg}) with $\Theta=\alpha$ and $\Theta=\mu\mu$ show the same time dependence, a monoexponential decay with the decay rate of the second Legendre polynomial. This is analogous to the step field-off-response discussed in Ref.\cite{cole}. The maximum position in $\lg P_2^\Theta(\omega)\rg$ therefore is located at frequency $\omega_{max}=6D_i$ for both terms.

The amplitude functions $A^\Theta(\Omega_p,D_i,t_p)$ are given by \cite{meinsenf}
\Be\label{respeineralal}
A^{\alpha}(\Omega_p,D_i,t_p) & = & \frac{1}{30} \beta \alpha E^2   \frac{\Omega_p}{D_i} \varepsilon ''(6D_i,2\Omega_p) \left[1-e^{-6 D_i t_p} \right]   \\ \nonumber
A^{\mu \mu}(\Omega_p,D_i,t_p) & = & \frac{1}{30} \beta^2 \mu^2 E^2  \varepsilon ''(2 D_i,\Omega_p) \times  \\
\label{respeineralmumu} & \hspace{-0.6cm} \times & \hspace{-0.6cm} \left( 3\varepsilon ''(4D_i,\Omega_p) \left[e^{-6 D_i t_p}-e^{-2 D_i t_p} \right] +\frac{5}{3} \varepsilon ''(6D_i,2\Omega_p) \left[ 1-e^{-6 D_i t_p} \right] \right)  
\Ee
with the dielectric loss defined as $\varepsilon ''(x,y)=xy/(x^2+y^2)$. In the context of frequency-selectivity the behavior of the amplitudes as a function of pump frequency is important. $A^{\mu \mu}(\Omega_p,D_i,t_p)$ plotted versus pump frequency shows a peak because the amplitudes become large if the arguments in the dielectric loss functions $\varepsilon ''(x,y)$ are equal, that means for $\Omega_p\approx 2D_i,3D_i,4D_i$. If for a moment we neglect the fact that the arguments of the dielectric loss functions have varying prefactors, then $E^2 \varepsilon ''(D_i,\Omega_p)^2$ represents the pump frequency-dependent energy absorbed by the system.

No frequency-selective behavior can be expected from the quasilinear $\alpha \alpha$-term, where the additional factor $\Omega_p$ appearing in the nominator of $A^{\alpha}(\Omega_p,D_i,t_p)$ does not allow for a 'resonant' energy absorption. This is because the product $\Omega_p \varepsilon ''(6D_i,2\Omega_p)$ is a strictly increasing function of $\Omega_p$ in contrast to $\varepsilon ''(6D_i,2\Omega_p)$ which exhibits a peak at $\Omega_p=3D_i$.

\subsection*{Heterogeneous model}
In a first discussion of Kerr effect responses in the heterogeneous model we neglect exchange effects. We choose the rotational diffusion constants distributed according to a generalized Gamma distribution with parameters such that the resulting (normalized) $(L=1)$-correlation has the form $\exp\left[ -(t/\tau)^{\beta_K} \right]$ with the KWW-exponent $\beta_K=0.5$ and $\tau =1$. The distribution of rotational diffusion constants is shown in the upper panel of Figure $1$. The lower panel shows the Kerr effect signal $\lg P_2^{\mu\mu}(\omega)\rg$ for different pump frequencies $\Omega_p=0.01,0.1,1,10,100$ (indicated by letters A,B,C,D,E in the figure). One cycle has been applied to the system for all pump frequencies. 

In order to understand the curves in Figure $1$ we refer to the previous section where we discussed the response of a single $D_i$. Here, the energy absorption was shown to have a maximum for $\Omega_p \approx 2D_i, 3D_i, 4D_i$. This means that in a distribution of $D_i$ most energy is absorbed if $D_i \approx \Omega_p/2$ holds. Thus, a pump frequency dependent selection is possible. The $D_i$ corresponding to $\Omega_p=10^{-2}\dots 10^2$ are marked by the vertical arrows in the distribution function (upper panel of Fig.$1$). Under the assumption that the response is mostly governed by the selected $D_i$ we obtain a maximum position at $\omega_{max}=6D_i$, thus at $\omega_{max}\approx 3 \Omega_p$. Although the prefactor $3$ is not exactly recovered in the extremum positions in the lower panel of Fig.$1$, we indeed observe a proportionality of extremum position (indicated by dotted lines) and pump frequency for $\Omega_p=100\dots 0.1$ (E,D,C,B). These shifts of the extremum position are the hallmark of dynamic heterogeneity. For $\Omega_p=0.01$ (A) the relative shift becomes weaker. Here, a selection of $D_i\approx 0.01/2$ does not determine the total response because these small $D_i$ are underrepresented in the distribution function, see the arrow (A) in the upper panel. In that case, rotational diffusion constants around $D_i \approx 0.05$ (B in the upper panel) determine the extremum position. This also holds for still smaller pump frequencies. Here, the extremum position becomes independent of $\Omega_p$ while the amplitudes of the signal strongly decrease. A similar behavior is observed at the high-frequency end of the distribution, which we do not show explicitly. The maximum position as a function of $\Omega_p$ will be discussed in detail later compared to results in the homogeneous limit.

In order to discuss the width of the responses, we added a Debye function (dotted line near C in the lower panel of Figure $1$). The heterogeneous response (C) is somewhat broadened in comparison. The curve for $\Omega_p=100$ (E) has a negative onset on the low-frequency side of the peak. Two remarks concerning this finding of a partially negative response function are appropriate. First, the negative sign of the response function is due to the first term in Eq.(\ref{respeineralmumu}), which is proportional to the difference $\left[\exp(-6 D_i t_p)-\exp(-2 D_i t_p) \right]$. This is negative for finite $t_p$ in contrast to the other term $\propto \left[1-\exp(-6 D_i t_p) \right]$. A negative response function therefore is a switch-on effect, since the former contribution vanishes for large $t_p$. Indeed, if several cycles are applied, all response functions become positive in the whole frequency range. Second, even for finite $t_p$, the negative term can be considered as an artefact of the rotational diffusion model. This is because the rates $6D_i$, $2D_i$ in the exponents are the decay rates of the second and the first Legendre polynomial, which are known to be of similar magnitude in real systems as discussed in Section II. In any model where these rates are very similar (e.g. the exchange model discussed in Section II) or equal (e.g. isotropic rotational jumps instead of rotational diffusion) no negative sign of the response occurs for arbitrary $t_p$. 

Figure $2$ shows the contributions of the quasilinear term $\lg P_2^\alpha(\omega)\rg$  with the same parameters as in Figure $1$. The extremum position is not proportional to pump frequency although it weakly depends on $\Omega_p$. While the pump frequency spreads over five decades, the maximum position shifts only for less than one decade. The shift of the extremum positions is further diminished if several cycles are applied. This demonstrates the expected result that the quasilinear term is not able to detect dynamic heterogeneities. In the following discussion, we will therefore focus on the $\alpha \mu\mu$-contribution. The quasilinear term can be viewed as a background contribution to the response function. As it is of quadratic order in the applied field like the $\alpha \mu \mu$-contribution, no phase-cycling procedure allows the extraction of either of the contributions. However, the terms should be easily distinguishable since the quasilinear signal is at a more or less constant position. Furthermore, the problem with the quasilinear term is trivially minimized if systems with large permanent dipole moment $\mu$ (compared to their polarizability $\alpha$) are studied. In this case, $\lg P_2^{\alpha} (\omega)\rg$ is much smaller than $\lg P_2^{\mu \mu} (\omega)\rg$.

We now discuss the influence of exchange on the Kerr effect response in the framework of the exchange model outlined in Section II. For our calculations we assume that the exchange rates factor into terms that depend on initial and final environment only, $\gamma_{ij}=\rho(\eps_i) X \exp(\beta \eps_j)$. Here, $\rho(\eps_i)$ (probability to end in final state $\eps_i$) is some normalized density of states and $X \exp(\beta \eps_j)$ (probability to leave initial state $\eps_j$) can be interpreted as thermally activated jumps out of environment $\eps_j$. Here, we have identified the so-far not specified $\eps_j$ with energies. The prefactor $X$ allows to adjust the 'strength' of exchange. We furthermore relate the rotational diffusion constants to $\eps_i$ as $D_i=Y \exp(\beta \eps_i)$.  The model is designed such that the total rate out of environment $\eps_i$ is proportional to the appropriate rotational diffusion constant, $\gamma_i\approx (X/Y) D_i$, meaning that slow environments exchange slowly and fast environments exchange fast. The ratio of $\gamma_i$ and $D_i$, given by $(X/Y)$, determines the influence of exchange on the decay of rotational correlation functions.

For the curves shown in Figure $3$ we chose $\rho(\eps_i)$ to be Gaussian with width $\sigma=3$ and we have set $\beta=1$. In order to obtain correlation times around unity we set $Y=100$. The upper panel of Figure $3$ shows normalized correlation functions for $L=1,2$ without exchange ($X/Y=0$, dashed lines) and in the presence of exchange ($X/Y=10$, solid lines). In the latter case the decay is strongly affected by exchange. All correlation functions shown can be well fitted with a function of KWW-type with a stretching exponent $\beta_K \approx 0.5\ $. The ratios of correlation times $\tau_L$ obtained from the fit are $\tau_1/\tau_2 \approx 3.0$ without exchange and $\tau_1/\tau_2\approx 1.3$ with exchange.

The nonlinear Kerr effect response $\lg P_2^{\mu\mu}(\omega)\rg$ in the presence of exchange is shown in the lower panel of Figure $3$. All parameters describing exchange are chosen as for the solid lines in the upper panel. In the limit without exchange, $(X/Y)=0$, the model discussed in Figure $1$ is analytically recovered with a modified distribution of rotational diffusion constants. The resulting curves in that case are very similar to the responses in Fig.$1$. For that reason we show only the curves with exchange ($X/Y=10$). The pump frequencies are chosen as $\Omega_p=10^{-2} \dots 10^{2}$ with one applied cycle for each frequency. The shifts of the extremum position obtained are very similar as in the heterogeneous case without exchange, cf. Fig.$1$. The peaks become slightly broader in the presence of exchange. No negative sign of the response occurs in any of the curves. 

We therefore conclude that exchange does not affect the possibility to detect dynamic heterogeneities. It is important to stress that the Kerr effect experiment does not allow to conduct information regarding the lifetime of dynamic heterogeneities.

\subsection*{Homogeneous model}
We now discuss the responses in a homogeneous scenario, where the distribution of rotational diffusion constants is replaced by a single, but time-dependent rotational diffusion constant. The time dependence is chosen in a manner that the correlation function for $L=1$ is proportional to $\exp\left[ -(t/\tau)^{\beta_K} \right]$ with $\beta_K=0.5$ and $\tau=1$. The correlation function is thus the same as in the situation we discussed in Figure $1$. Inserting Eqns.(\ref{glvont}),(\ref{homd}) into (\ref{almumu}) we find
\Be 
\nonumber & & \hspace{-1.6cm} \lg P_2^{\mu \mu}(t) \rg=\frac{1}{5} \beta^2 \mu^2 E^2 \left(\frac{\beta_K}{\tau}\right)^2  \exp\left[ -3 \left(\frac{t+t_p}{\tau} \right)^{\beta_K} \right] \times  \\ \label{homoalmumu}
& & \hspace{-1.6cm} \times \int_0^{t_p} dt' \int_0^{t'} dt'' \sin(\Omega_p t') \sin(\Omega_p t'') \left(\frac{t'}{\tau} \right)^{\beta_K-1} \left(\frac{t''}{\tau} \right)^{\beta_K-1}  \exp\left[ 2 \left(\frac{t'}{\tau} \right)^{\beta_K}\right]  \exp\left[ \left(\frac{t''}{\tau} \right)^{\beta_K}\right] .  
\Ee
The double integral does not depend on time $t$ and therefore represents only a prefactor to the response. The time-dependence is solely governed by the function $ \exp \left[ -3 \left(\frac{t+t_p}{\tau} \right)^{\beta_K} \right]$. Thus, the maximum position in $\lg P_2^{\mu\mu}(\omega)\rg$ is determined by the Fourier-transform of this function.

Figure $4$ shows the homogeneous responses for the same pump frequencies as in Figure $1$. One cycle has been applied for each pump frequency, $t_p=2\pi/\Omega_p$. It is this pump frequency-dependence of $t_p$ which leads to the shifts in the extremum positions in Figure $4$ (see the function above). Although the extrema are not at a constant position as usually expected in a homogeneous scenario, the shifts are less pronounced than in the heterogenous limit. The dashed curves corresponding to $\Omega_p=1$ and $\Omega_p=10$ are plotted negative. This is an example for the fact that signs of nonlinear responses are in general not easy to predict. The change of sign is not a systematic feature in our calculations.

\subsection*{Comparison of heterogeneous and homogeneous models}

In Figure $5$ we compare the maximum position in heterogeneous (upper straight line) and homogeneous (lower straight line) models, plotted versus pump frequency $\Omega_p$ for one applied cycle. The distribution in the heterogeneous case is as in Fig.$1$, and the time-dependent rotational diffusion constant of the homogeneous model is as in Fig.$4$. In the homogeneous case a maximum position independent of pump frequency is obtained only for sufficiently large $\Omega_p$. Clear shifts of the extremum position are observed in the heterogeneous limit. The shifting becomes weaker for $\Omega_p < 0.1$. This is due to the shape of the chosen distribution function as elucidated in the discussion of Fig.$1$. 
 
For a distinction between homogeneity and heterogeneity, the straight lines in Figure $5$ do not look too promising at first glance because the homogeneous extrema are not at a constant position. However, this finding should not be overemphasized. The observation of shifts in the homogeneous extremum positions relies directly on the lack of time-translational invariance of the specific homogeneous model used in the calculation. This leads to the shifted time $(t+t_p)$ in the function $\exp \left[ -3 \left(\frac{t+t_p}{\tau} \right)^{\beta_K} \right] $. The fact that the extrema are not at a constant position is only due to the appearance of the pump time $t_p$ in this function, which is related to the pump frequency via $t_p=2\pi/\Omega_p$. It is furthermore easy to eliminate this effect. If $t_p$ is chosen independent of the pump frequency, then the homogeneous model obviously yields extremum positions completely independent of the pump frequency. This can be achieved by applying several cycles $N(\Omega_p)\propto \Omega_p$ because of $t_p =2\pi N(\Omega_p)/\Omega_p$.
  
We added some extremum positions for that case in Fig.$5$, denoted by symbols. We chose $t_p=2\pi/0.01$, meaning that one cycle is applied for the smallest pump frequency considered, $\Omega_p=0.01$, and $10^4$ cycles are applied for the largest $\Omega_p=100$. Then the homogeneous model extrema (open symbols) do not depend on $\Omega_p$. The influence of several applied cycles on the extremum positions in the heterogeneous case is small. The extrema are located at somewhat smaller frequencies in comparison to the curve for one applied cycle. This is due to the pump time dependence of the function $A^{\mu \mu}(\Omega_p,D_i,t_p)$. 

The theoretical finding that the homogeneous ansatz leads to strictly constant extremum positions only if the pump time is constant for all applied frequencies also holds for NHB. If NHB is carried out as outlined in Ref.\cite{mynhb} and the dynamics are treated in a rotational diffusion ansatz as in this work, we find in the homogeneous limit the same expression as Eq.(\ref{homoalmumu}) up to prefactors and a modified time dependence. It is again the lack of time-translational invariance that leads to terms like $(t+t_p)^{\beta_K}$ in the appearing exponential functions. We therefore suggest also in NHB to apply several cycles for higher pump frequencies. Also the heterogeneous model for NHB leads to very similar results as presented in the previous sections. Shifts of the extremum position are observed if the pump frequencies are varied in the range defined by the distribution of rotational diffusion constants. 

\subsection*{Distribution width in the heterogeneous model}

Finally, we discuss the influence of the width of a distribution of $D_i$ in a heterogeneous model. A linear behavior of the extremum position as a function of pump frequencies, $\omega_{max} \propto \Omega_p^\eta$ with $\eta =1$, is obtained only if the distribution of rotational diffusion constants is sufficiently flat in the studied range. If the distribution is narrow, then usually exponents $\eta <1$ are found.

In Figure $6$, extremum positions obtained from the heterogeneous model with a logarithmic Gaussian distribution of rotational diffusion constants $g(D_i)\propto \exp\left(-[\ln D_i - \ln D_{\infty}]^2/[2\sigma^2] \right)$ are shown for three different distribution widths $\sigma=1.8$ (diamonds), $\sigma=1.2$ (squares) and $\sigma=0.6$ (circles). Exchange effects are neglected. In each case, $D_{\infty}$ was chosen such that the correlation time of the $(L=1)$-correlation function in a KWW-fit is $\tau_1=1$. The corresponding KWW-exponents obtained from the fits of the correlation function are $\beta_K=0.51$ (diamonds), $\beta_K=0.67$ (squares) and $\beta_K=0.88$ (circles). We applied the field for fixed pump time $t_p=2\pi/0.001$ independent of the pump frequency.

Heterogeneities are observed via shifts of the extremum position. Constant behavior of $\omega_{max}(\Omega_p)$ is observed for small and large $\Omega_p$. This is because, similar as in Figure $5$, the 'selected' $D_i$ in these cases are located beyond the main contributions to the distribution functions. In Fig.$5$ a constant behavior for large $\Omega_p$ is not reached in the frequency range considered due to the tailing of the corresponding distribution of rotational diffusion constants, see the upper panel of Fig.$1$. We do not expect that a constant extremum position for large $\Omega_p$ would be obtained experimentally. Here, other dynamics than isotropic reorientations will occur in real systems.

One can see that the rise of $\omega_{max}$ with $\Omega_p$ is the steeper the broader the distribution is, approaching the limit $\eta =1$ for a very flat distribution. Acceptable power law fits for the shown curves are obtained in the window $\Omega_p=0.1 \dots 10$. Here, we find the exponents $\eta \approx 0.84$ (diamonds), $\eta \approx 0.63$ (squares) and $\eta \approx 0.2$ (circles). 

\section*{IV. Conclusions}

We have calculated the Kerr after-effect response to an external oscillating electric field for reorientational dynamics in an isotropic system. Such an experiment is very similar to NHB, thus aiming at the question of the nature of the dynamics. The different assumptions of heterogeneous and homogeneous dynamics have been treated in a rotational diffusion model. We find that homogeneous and heterogeneous dynamics show different behaviour in the Kerr effect response function. 

We propose the Kerr effect as an alternative to NHB. NHB and Kerr effect have in common that they are both nonlinear experiments. The advantage of the Kerr effect is that the signal has no linear response background. In an experiment, one would measure the reequilibration of the polarizability after the sample has been excited by a sinusoidal electric field. The frequency of the latter is then varied. If the time-dependent functions obtained in this way are plotted in a Fourier-transformed representation (imaginary part), then extremum positions as a function of the pump frequency shift under the assumption of heterogeneous dynamics, and they are constant in the homogeneous limit. The Kerr effect signal is thus able to distinguish between dynamic heterogeneity and homogeneity.

The interesting relation between dynamic and spatial heterogeneities is not resolved yet. We therefore emphasize that the Kerr effect experiment outlined in this article does not allow to obtain any information about spatial aspects of the heterogeneities.

The influence of exchange, and thus a finite lifetime of heterogeneities on the responses is small. Exchange in our model does not affect the possibility of frequency selective modification.

It turned out that the general trends of homogeneous and heterogeneous limits are obtained more clearly if several cycles of the pump field are applied. This is because switch-on effects are minimized for longer pump times. In addition, the extremum positions in our homogeneous model depend on the pump time $t_p=2\pi N/\Omega_p$, and therefore on the pump frequency. In order to suppress this artificial dependence, we suggest to keep the pump time constant rather than the number of cycles applied if measurements with different $\Omega_p$ are compared. If, e.g., $N$ cycles (where ideally $N>1$) are applied for pump frequency $\Omega_p$, then $2N$ cycles should be applied if the pump frequency is $2\Omega_p$.

If the width of a relaxation time distribution decreases then the shifts of the extremum position with pump frequency become weaker. This point might be of interest in the investigation of a sample at different temperatures. A steeper rise of the extremum position with pump frequency for lower sample temperatures would indicate an increasing width of the corresponding distribution function. A point we have not discussed explicitly is that of course also combinations of heterogeneous and homogeneous behavior may occur in real systems. If the weight of homogeneous and heterogenous character of the dynamics changes with temperature this should be observable in more or less pronounced shifts of the extrema.

\section*{Acknowledgement}
This work has been supported by the DFG under contract No. Di693/1-2.

\newpage
\subsection*{Figure captions}
\begin{description}
\item[Fig.1 : ] Upper panel: The generalized Gamma distribution plotted as a function of rotational diffusion constants $D_i=(2\tau_D)^{-1}$, where $\tau_D$ are the more common relaxation times. Parameters are chosen such that the correlation function $C^1(t)$ has the form $\exp(-(t/\tau)^{\beta_K})$ with $\tau=1$ and $\beta_K=0.5\ $. \\ Lower panel: Imaginary part of the Fourier transformed Kerr effect response (contribution of order $\alpha \mu \mu$) in the heterogeneous model with the distribution of rotational diffusion constants shown in the upper panel. The pump frequencies are $\Omega_p=10^{n}$ with $n=-2,-1,0,1,2$ (left to right). One cycle of each frequency has been applied to the system. The prefactors $\beta^2\mu^2 E^2$ were set to unity. For details see the text.
\item[Fig.2 : ] As in the lower panel of Figure $1$ for the Kerr effect contribution of order $\alpha \alpha$. Pump frequencies are $\Omega_p=0.01\dots 100$ (extrema from left to right) with one applied cycle. Here, $\beta \alpha E^2$ was set to unity. 
\item[Fig.3 : ] Upper panel: Normalized correlation functions with exchange (solid lines) and without exchange (dashed lines) for the values of $L=1$ and $L=2$ in both cases, plotted versus time. The respectively faster decaying curve is for $L=2$, the slower decaying curve for $L=1$. \\Lower panel: Imaginary part of the Fourier transformed Kerr effect response (contribution of order $\alpha \mu \mu$) in the exchange model, with parameters as for the solid lines in the upper panel. Pump frequencies are $\Omega_p=0.01\dots 100$ (extrema from left to right) with one applied cycle. Here, $\beta^2\mu^2 E^2=1$. For details see text.   
\item[Fig.4 : ] Imaginary part of the Fourier transformed Kerr effect response (contribution of order $\alpha \mu \mu$) in the homogeneous model. Pump frequencies are $\Omega_p=0.01\dots 100$ (extrema from left to right) with one applied cycle. Dotted lines are plotted negative. We set $\beta^2 \mu^2 E^2=1$.
\item[Fig.5 : ] Extremum positions in the imaginary part of the Fourier transformed Kerr effect response of order $\alpha \mu \mu$, plotted versus pump frequency. The lines are for one applied cycle of each pump frequency in heterogeneous (upper solid line) and homogeneous cases (lower solid line). Connected symbols show the extremum position in the case of fixed pump time $t_p=2\pi/0.01$ (full symbols: heterogeneous, open symbols: homogeneous). For details see text.
\item[Fig.6 : ] Extremum positions in the imaginary part of the Fourier transformed Kerr effect response of order $\alpha \mu \mu$, plotted versus pump frequency. The pump time is $t_p=2\pi/0.001$ for all pump frequencies. The three curves represent different widths of the distribution of rotational diffusion constants with the values $\sigma =0.6$ (circles), $\sigma =1.2$ (squares) and $\sigma =1.8$ (diamonds). For details see text.
\end{description}

\begin{center}

\begin{figure}
\includegraphics[width=17cm]{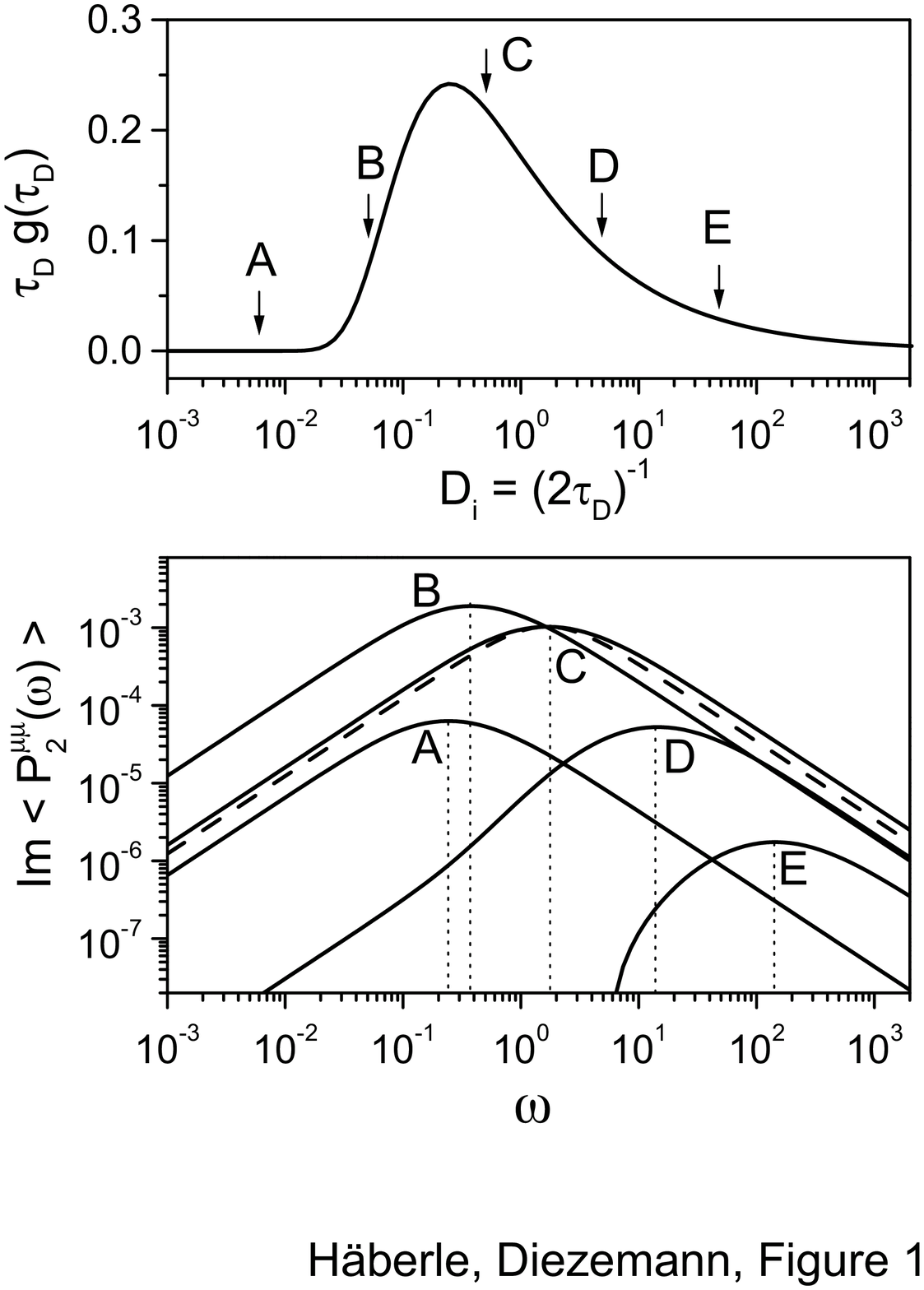}
\end{figure}

\begin{figure}
\includegraphics[width=17cm]{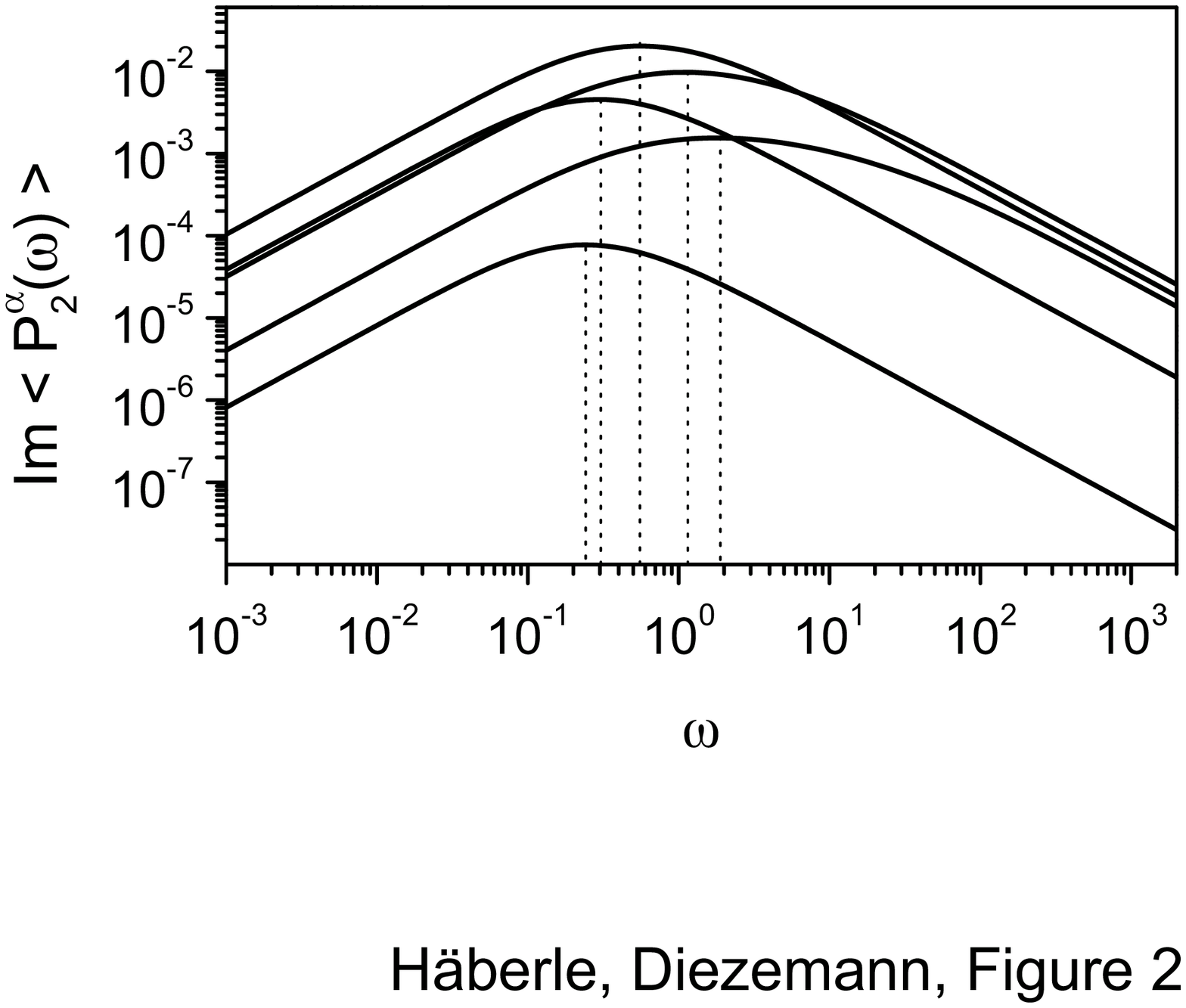}
\end{figure}

\begin{figure}
\includegraphics[width=17cm]{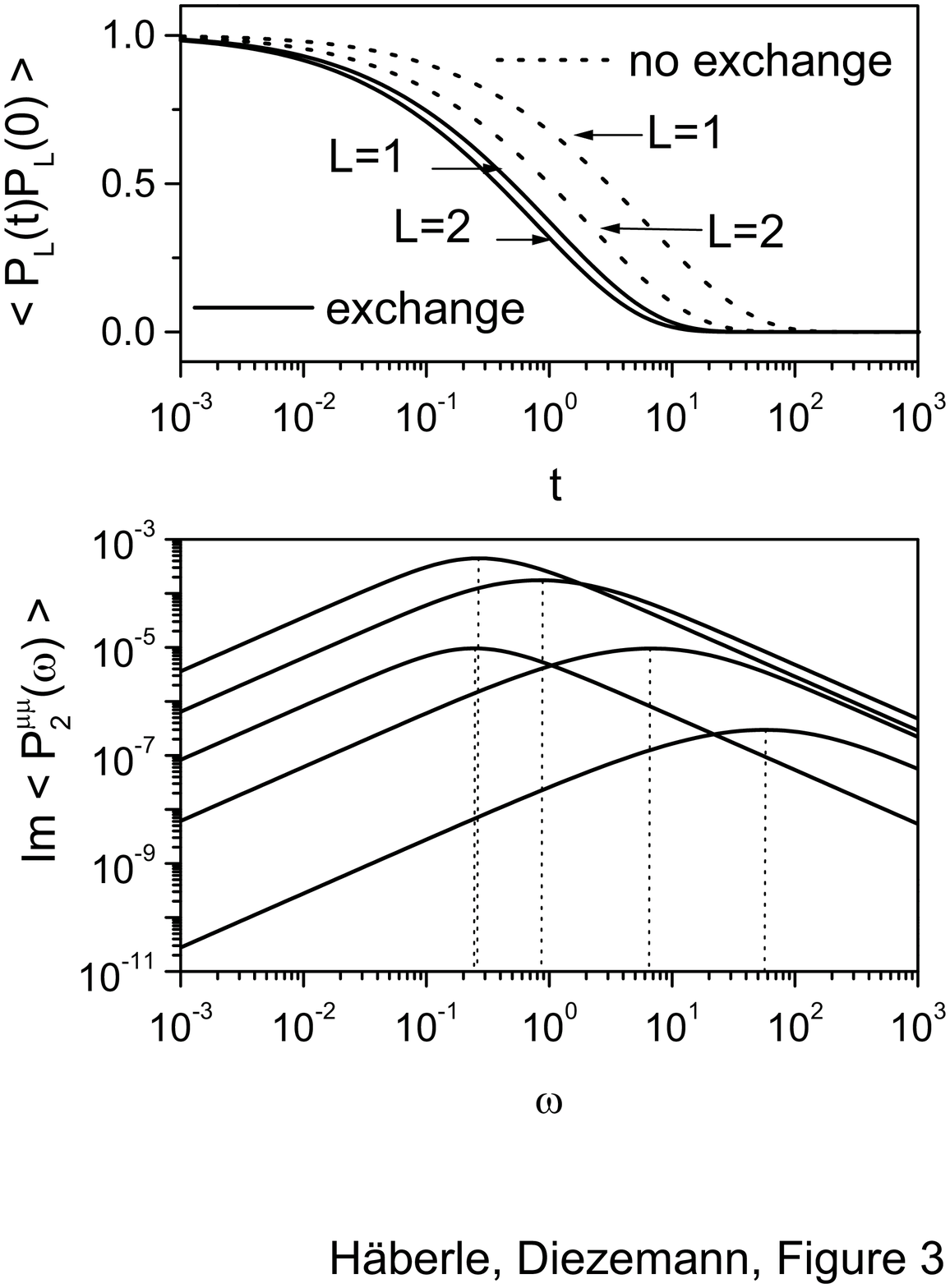}
\end{figure}

\begin{figure}
\includegraphics[width=17cm]{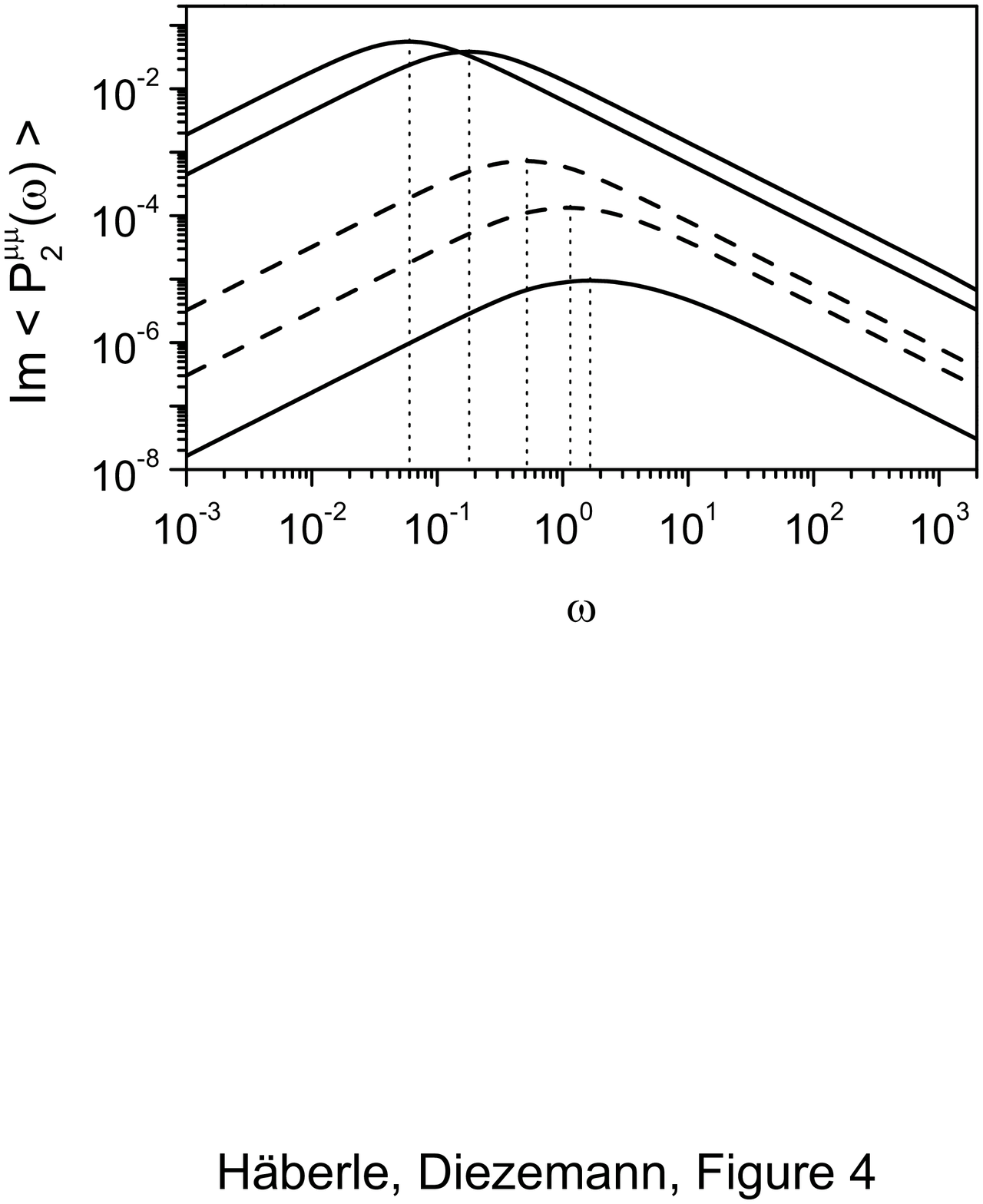}
\end{figure}

\begin{figure}
\includegraphics[width=17cm]{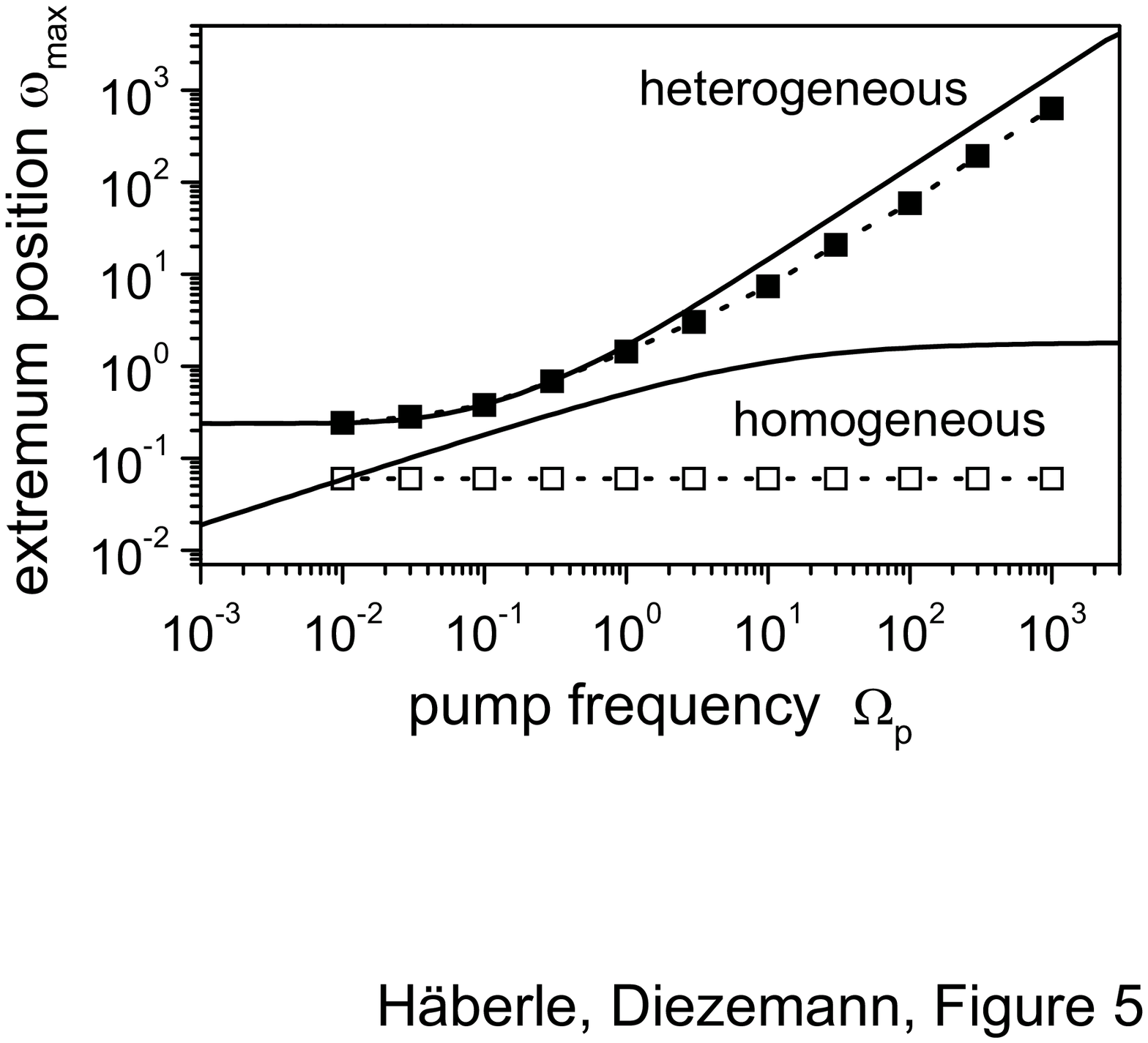}
\end{figure}

\begin{figure}
\includegraphics[width=17cm]{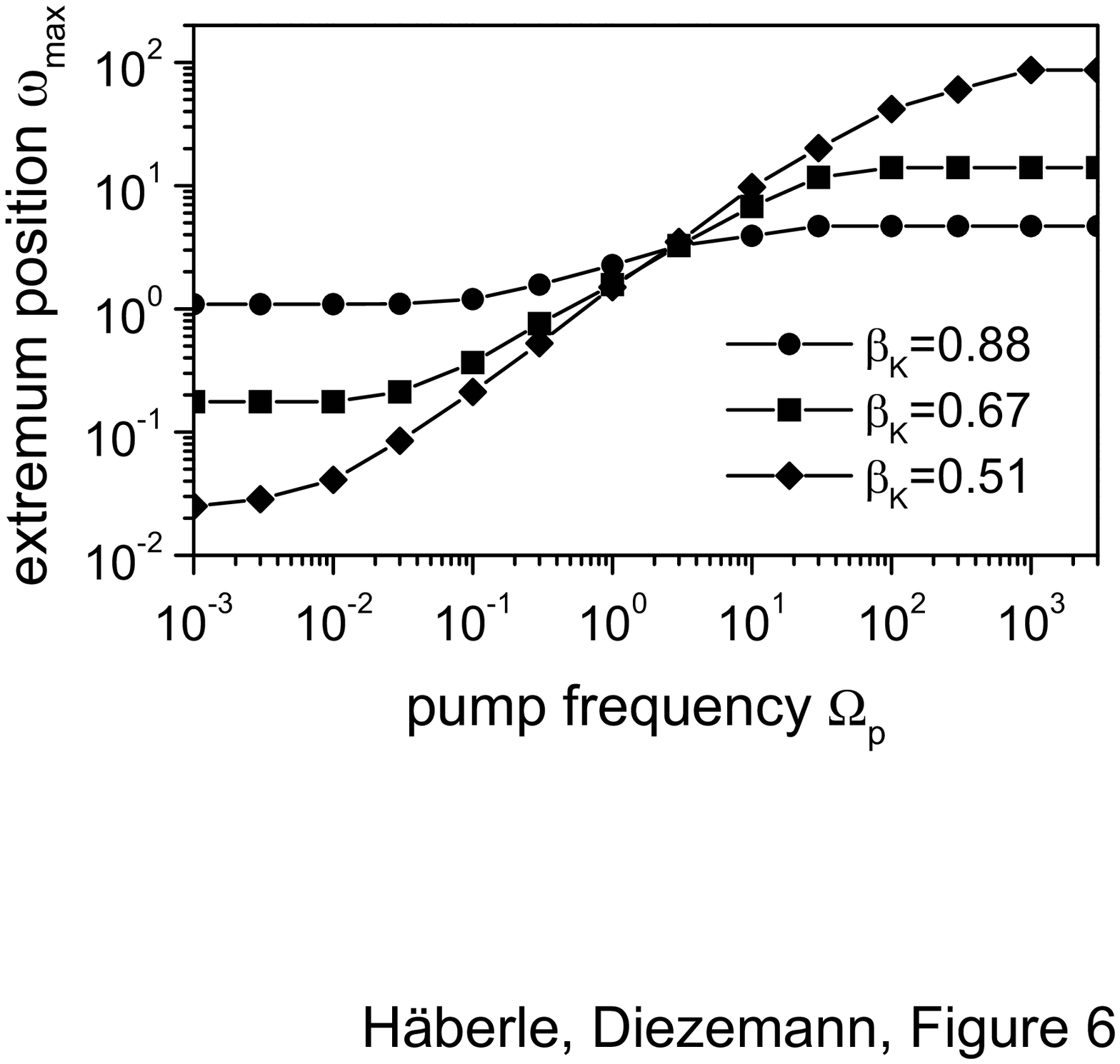}
\end{figure}

\end{center}


\begin{thebibliography}{XX}
\bibitem{meinsenf}
Expressions (\ref{respeinerallg}-\ref{respeineralmumu}) have been published in Ref.\cite{mykerr}.
\end{thebibliography}
\end{document}